\documentclass[12pt]{article}

\usepackage{amsfonts,bm}
\usepackage{amscd}
\usepackage{amsmath}

\usepackage{framed}
\usepackage{color}
\usepackage{slashed}
\usepackage{pdflscape}
\usepackage{enumerate}
\usepackage{multirow}
\usepackage[backref,pagebackref,linkcolor=blue]{hyperref}
\usepackage{mathrsfs}
\usepackage{comment}

\def\sss{\scriptscriptstyle}

\def\wt#1{\widetilde{#1}}

\def\sss{\scriptscriptstyle}

\def\be{\begin{equation}}
\def\ee{\end{equation}}
\def\beq{\begin{equation}}
\def\eeq{\end{equation}}
\def\bea{\begin{eqnarray}}
\def\eea{\end{eqnarray}} 
\def\beqa{\begin{equation}\begin{array}{l}}
\def\eeqa{\end{array}\end{equation}}
\def\eqn#1{(\ref{#1})}
\def\eqref#1{eq.~(\ref{eq:#1})}

 \def\G{{\it\Gamma}}

\def\nn{\nonumber}

\newcommand{\Rho}{{\mbox{\sf P}}}
\def\sideremark#1{\ifvmode\leavevmode\fi\vadjust{\vbox to0pt{\vss% the remark
 \hbox to 0pt{\hskip\hsize\hskip1em%                          will appear only
 \vbox{\hsize3cm\tiny\raggedright\pretolerance10000%          on the side
 \noindent #1\hfill}\hss}\vbox to8pt{\vfil}\vss}}}%
\newcommand{\one}{{\mathbb I}}

\begin{document}

\thispagestyle{empty}
\begin{flushright}
\framebox[.9\width]{\tiny \tt \begin{tabular}{c}CALT  68-2877\\ BRX-TH-658\end{tabular} }
\end{flushright}

\vspace{.8cm}
\setcounter{footnote}{0}

\begin{center}
{\Large{\bf 
Partial Masslessness and Conformal Gravity }
    }\\[10mm]

{\sc S. Deser$^\sharp$, E. Joung$^\flat$
and A. Waldron$^\natural$
\\[6mm]}

{\em\small  
$^\sharp$Lauritsen Lab,
Caltech,
Pasadena CA 91125
 and Physics Department, Brandeis University, Waltham,
MA 02454, 
USA\\ \href{mailto:deser@brandeis.edu}{\tt deser@brandeis.edu}}\\[5mm]
{\em\small  
$^\flat$ Scuola Normale Superiore and INFN,
Piazza dei Cavalieri 7, 56126 Pisa, Italy \\
\href{mailto:euihun.joung@sns.it}{\tt euihun.joung@sns.it}}\\[5mm]
{\em\small  
$^\natural$Department of Mathematics,
University of California, Davis, CA 95616,
USA\\ \href{mailto:wally@math.ucdavis.edu}{\tt wally@math.ucdavis.edu}}\\[5mm]

\bigskip

\bigskip

{\sc Abstract}\\
\end{center}

{\small
\begin{quote}

We use conformal, but ghostful, Weyl gravity to study its ghost-free, second derivative, partially massless (PM) spin 2 component in presence of  Einstein gravity with positive cosmological constant. 
Specifically, we consider both gravitational- and self- interactions of PM 
via the fully non-linear factorization of conformal gravity's Bach tensor into Einstein times Schouten operators.
We find that extending 
PM beyond linear order  suffers from familiar higher-spin consistency obstructions: 
it propagates only in Einstein backgrounds, and the conformal gravity route generates only the usual safe, Noether, cubic order vertices.

\bigskip

{\tt PACS: 04.62.+v, 04.50.-h, 04.50.Kd, 04.62.+v, 02.40.-k}

\bigskip

\end{quote}
}

\newpage
%\setcounter{page}{1}
%%%%%%%%%%%%%%%%%%%%%%%%%%%%%%%%%%%%%%%%%%%%%%%%%%%%%%%%%%%%%%%%%

%\vfill

%\tableofcontents

%\vfill

%\newpage

\section{Introduction and Review}
\label{IR}

Conformal, $d=4$, Weyl gravity (CG) provides a natural arena for studying 
the partially massless\footnote{Recall that  there are three varieties of spin 2 excitations in~dS: massive, massless and partially massless~\cite{Deser:1983tm,Deser:2001pe} 
The latter enjoy an interesting mixed behavior: In~dS they propagate
 lightlike, positive energy (inside the maximally accessible intrinsic dS horizon), helicity~$\pm 2,\pm1$ excitations in a unitary representation of the~dS isometry group~\cite{Higuchi:1986py,Deser:2001xr,Deser:2001wx,Deser:2003gw}. 
  This degree of freedom (DoF) count relies on the  gauge invariance~\eqn{pm_gauge} and a divergence constraint $\nabla^\mu \varphi_{\mu\nu}=\nabla_\nu \varphi$, where $\varphi:=\varphi^\rho_\rho$.
 } (PM) spin-2 field: 
 when expanded about de Sitter~(dS) backgrounds, its
kinematics consist of PM and the graviton~\cite{Maldacena:2011mk}.
Indeed, PM is invariant under the (tuned) sum of (linearized) diffeomorphism and  conformal transformations: 
 \begin{equation}\label{pm_gauge}
 \delta \varphi_{\mu\nu}=
 \big(\nabla_\mu\partial_\nu + \tfrac\Lambda 3\,g_{\mu\nu}\big)\alpha(x)\, .
 \end{equation}

In particular, we will use CG to  test whether the well-known gravitational- and self- coupling difficulties
encountered by higher-spin fields can be circumvented for PM, while avoiding the physically unacceptable ghost excitations CG shares with all higher derivative models. 
We will find first that while~PM can live in Einstein  (Einstein tensor proportional to the metric), rather than just dS (constant curvature) spaces~\cite{Deser:1983tm,Deser:2001pe,Dolan:2001ih}, 
it breaks down in more general classes of geometries. We then find that although PM's self-coupling
still faces the problems of higher-spin fields, in dS or Einstein backgrounds self-interacting cubic vertices can be 
obtained by contracting a Noether current with its corresponding gauge field. Consistency beyond cubic order is, however, difficult to maintain.
 CG avoids this pitfall at the cost of  ghost excitations, but its
 truncation to ghost-free~PM modes fails beyond its lowest, cubic,  order.

Interactions of~PM in four dimensions in particularly interesting because it is rigidly $SO(4,2)$ conformally invariant~\cite{Deser:2004ji}, just like its vector Maxwell counterpart. In fact PM can be
coupled to charged matter fields~\cite{Deser:2006zx} (see also~\cite{Zinoviev:2009hu}) which
suggests forming non-abelian multiplets--a still open problem. 
Instead, we will be concerned with the natural (singlet) tensor analogy and (abelian) self-interactions possibility. Results for cubic vertices derived from a 
St\"uckelberg approach were first given in~\cite{Zinoviev:2006im}
while a more general calculus of higher derivative~PM cubic vertices was developed in~\cite{Joung:2012rv}. Also, it has recently been suggested that a~PM limit of the putative massive gravity %of~\cite{deRham:2010kj} 
could be a candidate for an interacting~PM theory~\cite{deRham:2012kf}. 

Weyl transformations underlie PM's invariances, so CG is the obvious tool for studying its interactions.
While CG always has six excitations, the detailed spectra are background-dependent.
About flat space, it has two massless tensors and a photon with the same signature as one of them~\cite{2+2+2}, while in constant curvature backgrounds there is still a (cosmological) graviton, but now the (tensor+photon) combination becomes the PM mode with helicities $(\pm2,\pm1)$. In each case, the two sets of  modes are relatively ghost-like. The relative sign between PM's helicities depends on that of~$\Lambda$: In anti de Sitter (AdS), one can truncate the solution space to just
the unitary, massless graviton~\cite{Maldacena:2011mk,Lu:2011ks,Metsaev:2007fq} (for related analysis of higher derivative theories see~\cite{Lu:2011zk,Hyun:2011ej}).
The dS story is the  interesting one here because we can truncate, leaving either mode unitary;  the truncation to a unitary, PM mode being the relevant one.

A central ingredient of CG is  the  Yang--Mills  theory of the conformal group~$SO(4,2)$; it underlies an Ostrogradski
treatment in terms of a second derivative order  action~\cite{Kaku:1977pa} with field content directly corresponding to gravitons and~PM. It is also the
starting point for modern mathematical approaches to conformal geometries~\cite{BEG}.
Indeed, given the following gravity data: (i) the vierbein~$e_\mu{}^m$, (ii) the Levi--Civita connection~$\nabla_\mu$ and (iii)~the Schouten tensor
$\Rho_{\mu\nu}:=\frac12 (R_{\mu\nu}-\frac16\, g_{\mu\nu} R)$, then the   particular
$SO(4,2)$ Yang--Mills connection
$$
\nabla^{{\mathcal T}}_\mu= \one\, \nabla_\mu + \begin{pmatrix}0&-e_\mu{}^n & 0 \\ \Rho_\mu^m&0&e_\mu{}^m\\0&-\Rho_\mu^n&0\end{pmatrix}\, 
$$
plays a distinguished r\^ole. It is called the tractor connection~\cite{BEG}. {\it Any} connection~$\nabla$ faces three fundamental questions:
\begin{enumerate}
\item When does its curvature~$F=\nabla^2$ vanish?
\item When  are there parallel sections: vectors~$I$ such that~$\nabla I=0$?
\item When does~$F$ obey the Yang--Mills equations $\nabla^* F=0$? 
\end{enumerate}
The answers for the tractor connection~$\nabla^{\mathcal T}$ on a conformal 4-manifold~$(M,[g_{\mu\nu}])$ (one equipped with a conformal class of metrics~$[g_{\mu\nu}]=[\Omega^2 g_{\mu\nu}]$)~are:
\begin{enumerate}
\item When~$g_{\mu\nu}$ is conformally flat~\cite{BEG}.
\item When~$g_{\mu\nu}$ is conformal to an Einstein metric~\cite{BEG}.
\item When~$g_{\mu\nu}$ is Bach-flat~\cite{Goal} ({\it i.e.} a CG solution, see~\eqn{Bach}).
\end{enumerate}
While Einstein metrics are of prime importance, our main focus will be on Bach-flat metrics, since vanishing of the Bach tensor is the
 CG equation of motion that follows from the action\begin{eqnarray}\label{W2action}
S[g] = \tfrac18\int  \sqrt{-g}\ W^{\mu\nu\rho\sigma}\,W_{\mu\nu\rho\sigma} 
\, =\,  \tfrac14\int \sqrt{-g}\,\left( R^{\mu\nu}\,R_{\mu\nu}-{\tfrac13}\,R^2 \right),
\end{eqnarray}
modulo the Euler invariant.
%$:=\frac1{32\pi^2}\, \int  \sqrt{-g}\,\big(
%	R^{\mu\nu\rho\sigma}\,R_{\mu\nu\rho\sigma}
%	-4\,R^{\mu\nu}\,R_{\mu\nu}+R^{2}\big)$ is the integrated Gau\ss-Bonnet invariant.

The above three conditions  ascend in generality: conformally flat metrics are conformally Einstein and 
conformally Einstein metrics are Bach-flat.  A key question  is how to characterize Bach-flat metrics that are {\it not} conformally Einstein.
An answer involves~PM fields in an essential way: Consider the, symmetric, trace- and divergence-free Bach tensor
\begin{equation}\label{Bach}
B_{\mu\nu}:=-\Delta\,\Rho_{\mu\nu}+
\nabla^\rho\,\nabla_{(\mu}\,\Rho_{\nu)\rho}+W_{\rho\mu\nu\sigma}\,\Rho^{\rho\sigma}\, .
\end{equation}
which is invariant under local Weyl rescalings\footnote{In dimension three, the 2-index form  of the Weyl-invariant, traceless and conserved, Cotton tensor, $C_{\mu\nu}{}^m := \nabla_\mu \Rho_\nu{}^{\!m}-\nabla_\nu \Rho_\mu{}^{\!m}$ generates the``Weyl''-CS model~\cite{TMG}.} 
\begin{equation}\delta g_{\mu\nu} = 2\,\alpha\,g_{\mu\nu}
\label{rhoweyl}\quad
\Rightarrow\quad
\delta \Rho_{\mu\nu}=-\nabla_\mu\,\partial_\nu\,\alpha\, .
\end{equation}
Using the Bianchi identity $$\nabla^\nu \Rho_{\mu\nu} =\nabla_\mu \Rho\, ,\qquad \Rho:=\Rho^\rho_\rho\, ,$$ reordering covariant derivatives at the cost of terms quadratic in~$\Rho_{\mu\nu}$,  and defining the cosmological Schouten tensor
\begin{equation}\label{varphi}
\varphi_{\mu\nu}:= -\Rho_{\mu\nu}+\tfrac \Lambda 6\,g_{\mu\nu}\, ,
\end{equation}
the Bach tensor reads
\begin{eqnarray}\label{pm_eom}
B_{\mu\nu}(g,\varphi)
&=&\Delta\,\varphi_{\mu\nu}
- 2\,\nabla_{(\mu}\,\nabla^\rho\,\varphi_{\nu)\rho}+ g_{\mu\nu}\,\nabla^\rho\,\nabla^\sigma\, \varphi_{\rho\sigma}
+\nabla_\mu\,\nabla_\nu\,\varphi 
-g_{\mu\nu}\,\Delta\,\varphi\nonumber\\[2mm]
&&-\,2\,W_{\rho\mu\nu\sigma}\,\varphi^{\rho\sigma}
-\tfrac43\,{\Lambda}\,(\varphi_{\mu\nu} -\tfrac14\,g_{\mu\nu}\varphi) \, + \, O\big(\varphi^2\big)
%+4\varphi_{\mu\rho}\varphi^\rho_\nu - g_{\mu\nu}\varphi_{\rho\sigma}\varphi^{\rho\sigma}
\, . \label{BachPM}
\end{eqnarray}
If~$g_{\mu\nu}$ is close to an Einstein metric with cosmological constant~$\Lambda$, it follows that~$\Rho_{\mu\nu}\approx \frac\Lambda 6 g_{\mu\nu}$,
so~$\varphi_{\mu\nu}$ can be viewed a fluctuation and its quadratic terms  can, to leading order, be dropped. Upon setting~$g_{\mu\nu}$ to an Einstein metric,  the resulting linear equation for~$\varphi_{\mu\nu}$ is
exactly the~PM field equation in the Einstein background. In particular, observe that the Weyl transformation formula~\eqn{rhoweyl} for the Schouten tensor
yields the gauge invariance~\eqn{pm_gauge} of the~PM field~$\varphi_{\mu\nu}$:
the cosmological Schouten tensor corresponds to the~PM field in CG.

The fact that CG yields the PM equations in Einstein spaces motivates
 our two main questions:
\begin{enumerate}
\item Can~PM fields be coupled to gravitational backgrounds more general than Einstein?
\item Can CG be used as a catalyst to generate consistent, ghost-free, self-interactions for~PM fields, at least in Einstein backgrounds?
\end{enumerate}
Our answer to the first question is given in Section~\ref{GI}. There we review results from conformal geometry that completely characterize the failure of generic Bach-flat metrics (CG solutions) to be Einstein (cosmological Einstein gravity solutions). This gives the technology to study propagation of free~PM
fields in general backgrounds. We consider, then disprove the natural conjecture that Bach-flat backgrounds support consistent propagation of~PM fields.

In Section~\ref{SI} we review the truncation of CG  to cosmological Einstein gravity by turning off its~PM excitations. Then we consider the converse question: to what extent can the the configuration space of CG be split into~PM and 
graviton fields such that 
 the dynamical metric can be held to a fixed Einstein background, leaving a theory of interacting~PM fields? While this mechanism holds at linear
level for the field equations, we show it cannot be continued to non-linear interactions. This does not mean that no self-interacting~PM model exists, but rather that it could not be a CG-truncation. CG can still be used to generate cubic vertices that are consistent with 
the leading order~PM gauge invariance. While the construction of  cubic order vertices says nothing about the existence of a fully interacting theory, this   mechanism  generates them efficiently. 

In the Conclusion, we summarize our results and  discuss to what extent our CG based no-go statements extend to general~PM theories. We also speculate briefly about novel approximate cosmological solutions to Einstein gravity generated by a CG-dominated epoch, and also on higher-spin adaptations of our approach.

\section{PM-Gravitational Interactions}\label{GI}

A natural (but soon to be disproved!) conjecture is that Bach-flat backgrounds are the most general ones in which (linear)~PM fields propagate.
Striking evidence for this is the existence of Weyl invariant field equations enjoying double derivative gauge invariances precisely when the background is Bach-flat.
These models derive from ideas in conformal geometry: Let us return to the parallel condition~$\nabla^{\mathcal T} I = 0$ for the tractor connection defined in Section~\ref{IR}.
Denoting the so-called scale tractor by $I_M=(\rho,n_\mu,\sigma)$,  this condition becomes~\cite{BEG}
$$\left\{
\begin{array}{c}
\partial_\mu \sigma - n_\mu =0\, ,\\[1mm]
\nabla_\mu n_\nu + \sigma\, \Rho_{\mu\nu} + \rho \, g_{\mu\nu} = 0\, ,\\[1mm]
\partial_\mu\, \rho - \Rho_{\mu\nu}\,n^\nu  =0\, .
\end{array}\right.
$$
This system enjoys (local) conformal invariance; in particular, under local Weyl rescalings,~$\sigma \mapsto e^\alpha\, \sigma$, so setting~$\sigma=1$
is a legal gauge choice. Then the first equation gives~$n_\mu=0$,  the second equation states~$\Rho_{\mu\nu}\propto g_{\mu\nu}$: 
the parallel condition implies that the metric is (conformally) Einstein. (The converse is also true, see~\cite{BEG}). Returning to a general gauge and using the first and third
equations we can express the  scale tractor~$I_M$ in terms of~$\sigma$, as~$I_M=\big(-\frac14\,(\Delta + \Rho)\,\sigma, \nabla_\mu \sigma, \sigma\big)$.
The parallel conditions now reduce to 
$$
{\bm P}_{\mu\nu} \,\sigma = 0\, ,
$$
where the operator~${\bm P}_{\mu\nu}$, which maps scalars to trace-free symmetric tensors, is in fact conformally invariant. Explicitly
$$
{\bm P}_{\mu\nu}:= \nabla_{\{\mu}\,\partial_{\nu\}} + 
\Rho_{\{\mu\nu\}}\,,
$$
where $X_{\{\mu\nu\}}:=X_{(\mu\nu)}-\frac14\, g_{\mu\nu}\,X^\rho_\rho$
is the symmetrized trace-free part of $X_{\mu\nu}$, so  $g^{\mu\nu} X_{\{\mu\nu\}}=0$.
Zero modes of this operator correspond to conformally Einstein metrics; if~$\sigma\in\ker {\bm P}_{\mu\nu}$, then~$\sigma^{-2} g_{\mu\nu}$ 
is an Einstein metric~\cite{Gover:2006fa}. In addition, the gauge transformation of the trace-free components of a~PM field~$\varphi_{\mu\nu}$ are
generated by~${\bm P}_{\mu\nu}$. The key property, for our purposes, of the operator~${\bm P}_{\mu\nu}$ was discovered in~\cite{Gover:2006fa}:
 it permits the operator factorization of the Bach tensor as
\begin{equation}\label{detour_complex}
\!B_{\mu\nu}={\bm M}_{\mu\nu}^{\rho\sigma} \,{\bm P}_{\rho\sigma}\, ,\quad
{\bm M}^{\rho\sigma}_{\mu\nu}:=\delta^\rho_{\{\mu}\delta^{\sigma\phantom{\rho}}_{\nu\}}\Delta
-\delta_{\{\mu}^\rho \nabla^\sigma\, \nabla_{\nu\}}^{\phantom{\rho}}
-\tfrac13\, \delta_{\{\mu}^\rho \nabla_{\nu\}}^{\phantom{\rho}}\nabla^\sigma  -W^\rho{}_{\mu\nu}{}^\sigma\, .
\end{equation}
We observe that~${\bm M}_{\mu\nu}^{\rho\sigma}$ gives the non-linear answer to the  question posed in the Introduction: characterizing Bach-flat metrics
that are not conformally Einstein. We see that those require the range of~${\bm P}_{\mu\nu}$ to intersect the kernel of~${\bm M}_{\mu\nu}$; the operator ${\bm M}$ is also conformally invariant and maps trace-free symmetric tensors to trace-free symmetric tensors.

In mathematical terms, the identity~\eqn{detour_complex} means that  the sequence of three differential operators ${\bm P}_{\rho\sigma}, {\bm M}^{\rho\sigma}_{\mu\nu}, {\bm P}^\dagger{}^{\mu\nu}$ (the third being the adjoint of~${\bm P}_{\mu\nu}$)
form a complex whenever the metric is Bach-flat. In~\cite{Gover:2006fa}, this is termed a Yang--Mills detour complex. Physically, it implies that the field equation
\begin{equation}\label{ym_cmplx_eom}
{\bm M}^{\rho\sigma}_{\mu\nu}\,  \wt\varphi_{\rho\sigma} = \Delta\, \wt \varphi_{\mu\nu}-\nabla^\sigma\, \nabla_{\{\mu}\,\wt\varphi_{\nu\}\sigma} 
-\tfrac 13\, \nabla_{\{\mu}\,\nabla^\sigma\, \wt\varphi_{\nu\}\sigma}
-W^\rho{}_{\mu\nu}{}^\sigma\,\wt\varphi_{\rho\sigma}=0\, ,
\end{equation}
for a trace-free symmetric tensor~$\wt \varphi_{\mu\nu}=:\varphi_{\{\mu\nu\}}$, enjoys the double derivative gauge invariance (and associated double derivative Bianchi identity)
$$
\delta \wt\varphi_{\mu\nu} = {\bm P}_{\mu\nu} \alpha = \big(\nabla_{\{\mu}\nabla_{\nu\}} +\wt \Rho_{\mu\nu}\big) \alpha\, ,
$$
in Bach-flat backgrounds. This was the  motivation for  our original conjecture that~PM fields could propagate in them. We now proceed to  disprove it and give necessary consistency conditions for~PM-compatible backgrounds.

The Bach tensor, since it arises from a metric variational principle, is necessarily divergence-free, $\nabla^\mu {\bm M}_{\mu\nu}^{\rho\sigma}\, {\bm P}_{\rho\sigma}=0$. However, it is neither true that~$\nabla^\mu {\bm M}_{\mu\nu}^{\rho\sigma}=0$, nor that~$\nabla^\mu {\bm M}_{\mu\nu}^{\rho\sigma}=O(\nabla)$ (rather this operator is cubic in derivatives). But consistent PM propagation relies on a divergence constraint\footnote{The DoF count for~PM starts with  ten off-shell fields~$\varphi_{\mu\nu}$, minus four DoF thanks to the divergence  constraint~$\nabla^\mu \varphi_{\mu\nu}=\nabla_\nu \varphi$,
minus two further DoF by the scalar gauge invariance, yielding a total of four on-shell excitations.}; for 
 a~PM field equation (derived from an action) this requirement is precisely expressed by the condition~$\nabla^\mu {\bm M}_{\mu\nu}^{\rho\sigma}=O(\nabla)$.
 
The failure of the field equation~\eqn{ym_cmplx_eom} to imply an appropriate divergence constraint does not yet rule out~PM
fields interacting with backgrounds more general than Einstein spaces, because we may still enlarge the space of field equation  and
gauge operators,~${\bm M}_{\mu\nu}^{\rho\sigma}$ and~${\bm P}_{\mu\nu}$ respectively, by relaxing their trace-free and conformal invariance properties.
To test that we make the following generalization
\begin{eqnarray*}
{\bm M}^\prime{}_{\mu\nu}^{\rho\sigma}\!\!\!&=&{\bm G}_{\mu\nu}^{\rho\sigma}
-\big(\delta^\rho_{(\mu}\,\delta^{\sigma\phantom{\rho}}_{\nu)}-g_{\mu\nu}\,g^{\rho\sigma}\big)\, \Rho
+\alpha_1\,\delta^\rho_{(\mu}\,\wt\Rho{}_{\nu)}^{\sigma\phantom{\rho}} 
+\alpha_2\,\big(g_{\mu\nu}\,\wt\Rho{}^{\rho\sigma}+\wt\Rho_{\mu\nu}\,g^{\rho\sigma}\big)
\ ,\\[2mm]
{\bm P}^\prime_{\mu\nu}&=&\nabla_\mu\,\partial_\nu + \tfrac12\, \Rho\,  g_{\mu\nu}  + \beta\,\wt\Rho_{\mu\nu}\, ,
\end{eqnarray*}
where the cosmological Einstein operator 
\begin{eqnarray}
{\bm G}_{\mu\nu}^{\rho\sigma}&:=&\big(\delta^\rho_{(\mu}\delta^{\sigma\phantom{\rho}}_{\nu)}-g_{\mu\nu}g^{\rho\sigma}\big)\big(\Delta-\Rho\big)
-2\,\nabla_{(\mu} \nabla^\rho \delta^\sigma_{\nu)}+\nabla_{(\mu}\nabla_{\nu)} g^{\rho\sigma}+g_{\mu\nu}\nabla^\rho\nabla^\sigma
\nonumber\\[1mm]&&-\,2\,W^\rho{}_{\mu\nu}{}^\sigma
-8\,\widetilde\Rho{}_{\{\mu}^\rho\delta^{\sigma\phantom{\rho}}_{\nu\}}-\tfrac32\,  g_{\mu\nu} \Rho\,  g^{\rho\sigma}\, ,\label{Gop}
 \end{eqnarray}
 is identically conserved
~$$
 \nabla^\mu {\bm G}_{\mu\nu}^{\rho\sigma}=0\, ,
~$$
in Einstein backgrounds. The equation of motion of cosmological Einstein gravity linearized about an Einstein metric 
is~${\bm G}_{\mu\nu}^{\rho\sigma} \,\varphi_{\rho\sigma}=0$.

The above ansatz is the most general one obeying the following requirements:
\begin{enumerate}
\item The operators ~${\bm M}^\prime{}_{\mu\nu}^{\rho\sigma}$ and~${\bm P}^\prime_{\mu\nu}$ are second order in~$\nabla$ or derivatives on the
metric~$g_{\mu\nu}$.
\item The operator~${\bm M}^\prime{}_{\mu\nu}^{\rho\sigma}$ is self-adjoint,  to ensure the existence of an action principle.
\item The divergence~$\nabla^\mu {\bm M}^\prime{}_{\mu\nu}^{\rho\sigma}$ is an operator no more than linear in~$\nabla$, to ensure
that solutions of~${\bm M}^\prime{}_{\mu\nu}^{\rho\sigma}\,\varphi_{\mu\nu}=0$ obey a first order constraint.
\item The operator product~${\bm M}^\prime{}_{\mu\nu}^{\rho\sigma}{\bm P}^\prime_{\rho\sigma}$ vanishes when~$g_{\mu\nu}$ is
an Einstein metric; this fixes their leading terms to be operators corresponding to the linear~PM equation of motion~\eqn{pm_eom}
and its double derivative gauge invariance~\eqn{pm_gauge}. The remaining freedom in the ansatz therefore depends  only on the trace-free
Schouten tensor~$\wt \Rho_{\mu\nu}$, since that quantity vanishes for Einstein metrics.
\end{enumerate} 

It remains to compute the product 
${\bm M}^\prime{}_{\mu\nu}^{\rho\sigma}\,{\bm P}^\prime_{\rho\sigma}$. The result can be arranged as an expansion in the gradient 
operator~$\nabla$. By construction, terms of order~$\nabla^4$ and~$\nabla^3$ necessarily vanish. Prefactors of the terms order~$\nabla^2$ 
only involve~$\wt \Rho_{\mu\nu}$ which we are now assuming to be non-vanishing, since we wish to investigate metrics that are {\it not} Einstein:
we must choose the constants~$(\alpha_1,\alpha_2,\beta)$ accordingly and find
$$
\alpha_1=4+2\beta \quad\mbox{and}\quad \alpha_2=-\beta\, .
$$
The analysis of terms order~$\nabla$ and lower is more complicated. First we consider the trace~$g^{\mu\nu}{\bm M}^\prime{}_{\mu\nu}^{\rho\sigma}{\bm P}^\prime_{\rho\sigma}$ at order~$\nabla$ and find~$3\beta (\nabla_\rho \Rho)\nabla^\rho$. There are two possibilities, either~$\beta=0$ or the background metric has constant scalar curvature.
Since the latter  would rule out the PM conjecture in question,  we choose~$\beta=0$.
We then find~$g^{\mu\nu}{\bm M}^\prime{}_{\mu\nu}^{\rho\sigma}{\bm P}^\prime_{\rho\sigma}=-3(\Delta \Rho)$, which requires the scalar curvature to be harmonic,
and hence also rules out the conjecture.

Having disproved the conjecture,
one may still investigate whether some  background condition weaker  than Bach-flat, but still less stringent than  Einstein, could yield consistent propagation.
The terms remaining at order~$\nabla$ in~${\bm M}^\prime{}_{\mu\nu}^{\rho\sigma}{\bm P}^\prime_{\rho\sigma}$ are
$$
\beta g_{\mu\nu} (\nabla_\rho \Rho)\nabla^\rho-(\beta-2)(\nabla_{(\mu}\Rho)\nabla_{\nu)}
+2(\beta-1)(\nabla_\rho \Rho_{\mu\nu})\nabla^\rho-2\beta(\nabla_{(\mu}\Rho_{\nu)\rho})\nabla^\rho\, ;
$$
clearly no choice of~$\beta$ removes all of them. Instead, we can restrict the background, one option being to Ricci-symmetric spaces, defined by~$\nabla_\rho \Rho_{\mu\nu}$ $=$~$0$. This condition is weaker than  Einstein, but need not imply Bach-flat.
However, even then we must cancel all terms in~${\bm M}^\prime{}_{\mu\nu}^{\rho\sigma}{\bm P}^\prime_{\rho\sigma}$ of order~$\nabla^0$. In general backgrounds these are
\begin{eqnarray*}
-\beta\,B_{\mu\nu}+2\beta^2\,\Rho^\rho_{(\mu}\Rho^{\phantom{\rho}}_{\nu)\rho}-\tfrac12 (\beta-1)(\beta+3)\, \Rho \, \Rho_{\mu\nu}
-\tfrac12 (\beta-2)\, \nabla_\mu\partial_\nu \Rho\\
 +\, g_{\mu\nu}\big[\tfrac12(\beta-2)\,\Delta \Rho-\beta(\beta+1)\,\Rho_{\rho\sigma}\Rho^{\rho\sigma}
          +\tfrac18(\beta+2)(3\beta-2)\,\Rho^2
          \big]\, .
\end{eqnarray*}
Even for a Ricci-symmetric space,  no choice of~$\beta$ removes all remaining terms quadratic in the Schouten tensor and its trace.
We also see no strong physical motivation to single out backgrounds with covariantly constant Einstein tensor  subject to a further quadratic curvature constraint.

\section{Self-Interactions}\label{SI}

To study self-interactions, we must first  recast our derivation of the~PM field equation~\eqn{pm_eom} in terms of the CG action~\eqn{W2action}. The latter
can be rewritten  
as a two-derivative action by introducing an auxiliary field~$\varphi_{\mu\nu}$~\cite{Kaku:1977pa}:
\renewcommand\G{{\mathcal G}}
\be\label{kaku}
S[g,\varphi]=-\int\sqrt{-g}\,
\Big[ \tfrac{\Lambda}6\left(R-2\Lambda\right)
+\varphi^{\mu\nu}\,\G_{\mu\nu}
+\varphi^{\mu\nu}\,\varphi_{\mu\nu}-\varphi^2\Big]\,,
\ee
where~$\G_{\mu\nu}:= G_{\mu\nu}+\Lambda\,g_{\mu\nu}$ is 
the cosmological Einstein tensor. 
Upon completing the square, we see that  the auxiliary field becomes the cosmological Schouten tensor~\eqn{varphi}.
To analyze the spectrum of the theory about an Einstein background~$\bar g_{\mu\nu}$ with cosmological constant~$\Lambda$, we  
linearize  in metric perturbations~$h_{\mu\nu}=g_{\mu\nu}-\bar g_{\mu\nu}$. 
Keeping terms quadratic in fluctuations and making the field redefinition
\begin{equation}
h_{\mu\nu}\,\to\,h_{\mu\nu}+\tfrac\Lambda6\,\varphi_{\mu\nu}\,.
\label{lin_redef}\end{equation}
yields the action (the metrics appearing in ${\bm G}$ and ${\bm F}$  are set to $\bar g_{\mu\nu}$)
%(immediately dropping the superfluous primes on the new field variables~$h'$ and~$\varphi'$)
%\edz{A: Can we give the correct prefactor here and check sign o mass term}	
\begin{equation}\label{lin_act}
	S^{\sss\rm (2)}[h,\varphi]
	=\tfrac14\int\sqrt{-\bar g}\,  
	\Big[-\tfrac\Lambda6\,h^{\mu\nu}\,{\bm G}_{\mu\nu}^{\rho\sigma}h_{\rho\sigma}
	+\tfrac{6}\Lambda\,\varphi^{\mu\nu}
	\left({\bm G}_{\mu\nu}^{\rho\sigma}
	-\tfrac23\,{\Lambda}\,{\bm F}^{\rho\sigma}_{\mu\nu}\right)\varphi_{\rho\sigma}\Big]\, .
\end{equation}
Here~$-{\bm G}_{\mu\nu}^{\rho\sigma}\,h_{\rho\sigma}/2$ is the linearized 
%\edz{A: check sign of Einstein tensor}
cosmological Einstein tensor defined in~\eqn{Gop} and all indices are moved by~$\bar g_{\mu\nu}$.
The Pauli--Fierz (PF) mass operator is defined as
${\bm F}^{\rho\sigma}_{\mu\nu}:=\delta^\rho_{\mu}\,\delta^\sigma_{\nu}-
g_{\mu\nu}\, g^{\rho\sigma}$, so the~PM field equation is
$$\big({\bm G}_{\mu\nu}^{\rho\sigma}-\tfrac23\,{\Lambda}\,{\bm F}^{\rho\sigma}_{\mu\nu}\big)\,\varphi_{\rho\sigma}=0\, .$$ 
Thus, the first term of~\eqn{lin_act} is linearized Einstein--Hilbert,
while the terms with round brackets (the sum of the linearized gravity kinetic term and a  Pauli--Fierz mass term tuned
to the~PM value~$m^2=2\Lambda/3$) give the~PM theory, all in an Einstein background. 
Hence the model describes the ``difference'' of  massless and~PM excitations. 
Moreover, integrating out (at linear level) the field~$\varphi_{\mu\nu}$ appearing before the field redefinition~(\ref{lin_redef}), gives the fourth order equation
$${\bm B}_{\mu\nu}^{\rho\sigma}\,h_{\rho\sigma}=0\, ,\qquad\mbox{where}\qquad {\bm B}_{\mu\nu}^{\rho\sigma}:={\bm G}_{\mu\nu}^{\alpha\beta}\, {\bm F}^{-1}{}_{\alpha\beta}^{\gamma\delta}\, {\bm G}^{\rho\sigma}_{\gamma\delta}
-\tfrac23\,{\Lambda}\, {\bm G}_{\mu\nu}^{\rho\sigma}\, ,$$ 
for the original metric fluctuations. Indeed,~${\bm B}_{\mu\nu}^{\rho\sigma}\,h_{\rho\sigma}$
%\edz{A: Check coefficients} 
is the Bach tensor linearized about an Einstein background.

The relative sign
of the two parts of the linearized action~\eqn{lin_act} reflects the unavoidable relative ghost  structure. In particular, states with~$\varphi_{\mu\nu}=0$
constitute a unitary, massless spin~$s=2$ spectrum. When the cosmological constant is positive (dS), states with~$h_{\mu\nu}=0$ correspond to a unitary
PM  spectrum. We now proceed to study the latter truncation;
a key step is to understand the model's gauge structure. At linear level, the graviton~$h_{\mu\nu}$ enjoys a linearized diffeomorphism
symmetry\footnote{As an aside, we observe that the derivation of the linear PM model from Weyl invariant CG theory gives a novel proof of the $SO(4,2)$ conformal invariance of PM excitations. (In fact, conformal invariance was the original rationale behind the PM model~\cite{Deser:1983tm}, and is enjoyed by all maximal depth, four-dimensional PM theories of generic spin~\cite{Deser:2004ji}.) In detail, whenever a field is coupled to the metric, maintaining Weyl invariance, then setting the metric to a background yields an action that enjoys any conformal isometries as symmetries. Thus the non-linear model generated by setting the metric in~\eqn{kaku} to a background is guaranteed to enjoy this  symmetry; since it holds order by order in~$\varphi$, it is also a symmetry of  linearized~PM. }~$\delta h_{\mu\nu}=\nabla_{\mu}\,\xi_{\nu}+\nabla_{\nu}\,\xi_{\mu}$ 
while the~PM field~$\varphi_{\mu\nu}$ transforms according to the double derivative scalar variation~\eqn{pm_gauge}; at linear level each field is inert under the other's transformations.
In fact, the~PM gauge symmetry is inherited from the Weyl symmetry of CG.
The full non-linear action~\eqn{kaku} is invariant under both gauge transformations,
%\edz{A: Changed signs here to the standard ones!}
\be\label{nonlin tr}
	\delta g_{\mu\nu}=\nabla_{\mu}\,\xi_{\nu}+\nabla_{\nu}\,\xi_{\mu}+2\,\alpha\,g_{\mu\nu}\,,\qquad
	\delta \varphi_{\mu\nu}=
	{\mathcal L}_{\xi}\,\varphi_{\mu\nu}
	+\big(\nabla_{\mu}\,\partial_{\nu}+\tfrac\Lambda3\,g_{\mu\nu}\big)\,\alpha\, .
\ee	
The metric transformation is now a sum of diffeomorphism  and  Weyl transformations as is the~$\varphi_{\mu\nu}$ transformation:
${\mathcal L_{\xi}}$ is the Lie derivative along the vector field~$\xi$ and the Weyl term follows from the transformation of the Schouten tensor~\eqn{rhoweyl}. 

Without incurring the ghost problem of CG, 
 we may search for some combination of fields that, when held to an appropriate background, yields a consistent
truncation to a self-interacting~PM model.\footnote{Indeed, the converse version of this procedure can be applied to 
produce cosmological gravity from
CG for the full, non-linear theory:
Examining the gauge transformations~\eqn{nonlin tr}, we see that the~PM background~$\varphi_{\mu\nu}=0$
is preserved by diffeomorphisms but not Weyl transformations. Hence, setting~$\varphi_{\mu\nu}=0$ yields a diffeomorphism 
invariant theory; performing this substitution in the action~\eqn{kaku} yields cosmological Einstein gravity.}
We must now find the proper  combination of fields to set to a background that yields the desired decoupling.
At linear level, the answer to this requirement is given by the field redefinition~\eqn{lin_redef}. There,  the choice for the metric
fluctuations~$h_{\mu\nu}=0$ is respected by~PM gauge transformations. This substitution in the linearized action~\eqn{lin_act}
yields the free~PM action in an Einstein background.
Therefore we begin by positing a candidate for a
 non-linear version of the field redefinition~\eqn{lin_redef} (that mixes~$g_{\mu\nu}$ and~$\varphi_{\mu\nu}$) such that a consistent~PM theory results from holding the redefined metric
to a suitable fixed value:
\begin{equation}\label{field_redef}
\left\{
\begin{array}{ccl}
g_{\mu\nu}&\to&g_{\mu\nu}+\tfrac6\Lambda\,\varphi_{\mu\nu}+J_{\mu\nu}(g,\varphi)\\[2mm]
\varphi_{\mu\nu}&\to& \varphi_{\mu\nu}+K_{\mu\nu}(g,\varphi)\,.
\end{array}\right.
\end{equation}
We take~$J_{\mu\nu}$ and~$K_{\mu\nu}$ to start at second order in~$\varphi_{\mu\nu}$ so as to preserve the linear level choice~\eqn{lin_redef}.
With this field redefinition, the CG action~\eqn{kaku} reduces to that of a 
``matter'' field~$\varphi_{\mu\nu}$ coupled to a (dynamical) metric:
\be\label{Weyl as ordinary}
	S[g,\varphi]=\int \sqrt{-g}\,\big[
	-\tfrac{\Lambda}6\,(R-2\,\Lambda)
	+\tfrac{6}\Lambda\,\mathscr L_{\rm\sss~PM}(\varphi,\nabla\varphi)\,\big]\,,
\ee
where~$\mathscr L_{\rm\sss~PM}$ is the candidate~PM Lagrangian.
Its $\varphi_{\mu\nu}$ dependence  
is highly non-linear, with self-interactions coming from re-expressing all the original metric dependence of the action~\eqn{kaku}
in terms of the shifted combination~$g_{\mu\nu}+\frac6\Lambda\, \varphi_{\mu\nu}+J_{\mu\nu}$. After making this expansion, we set~$g_{\mu\nu}$ to 
any Einstein metric with cosmological constant~$\Lambda$. This leaves us with the~PM candidate
\begin{equation}
\label{pmaction}
S_{\rm \sss PM}[\varphi]=\tfrac{6}\Lambda\,
\int \sqrt{-g}\,\mathscr L_{\rm\sss~PM}(\varphi,\nabla\varphi)\, ,
\end{equation}
to be computed as an expansion in
$\varphi_{\mu\nu}$:
\be
	\mathscr L_{\rm\sss PM}
	=\tfrac14\,\varphi^{\mu\nu}\left({\bm G}_{\mu\nu}^{\rho\sigma}
	-\tfrac23\,{\Lambda}\,{\bm F}^{\rho\sigma}_{\mu\nu}\right)\varphi_{\rho\sigma}
	+	\sum_{n=3}^{\infty}\mathscr L_{\rm\sss~PM}^{\sss(n)}\,.
\ee
The absence of a term linear in~$\varphi_{\mu\nu}$  follows from the linearized analysis and relies on the fact that~$g_{\mu\nu}$ is now an Einstein metric. 
Furthermore, we notice that the arbitrariness of the field redefinition~\eqn{field_redef} encoded by~$J(\varphi)$ and~$K(\varphi)$ really 
amounts only to a field redefinition~$\varphi_{\mu\nu}\,\to\,\varphi_{\mu\nu}+J_{\mu\nu}(\varphi)$  in the candidate action~(\ref{pmaction}) when 
$J_{\mu\nu}=K_{\mu\nu}$. Moreover, examining the full non-linear form~\eqn{kaku}, we see that since~$K_{\mu\nu}$ is at least second order
in~$\varphi_{\mu\nu}$, it only contributes to vertices at least quartic in the~PM field. Hence, with the understanding that we only quote
results for vertices up to possible $\varphi_{\mu\nu}$ field redefinitions, we may set
$J_{\mu\nu}(\varphi)=0=K_{\mu\nu}(\varphi)$ without any loss of generality for our cubic order results.

Before presenting our explicit cubic vertices, let us show that there is no fully non-linear truncation of CG to an interacting 
PM theory.
(This neither annuls consistency of the cubic vertices with respect to linearized gauge 
transformations, nor rules out any other  ultimate theory of self-interacting~PM fields.) To determine whether a truncation that  
takes~$g_{\mu\nu}$ to be a fixed Einstein background is consistent, we must study the gauge invariances of the theory.
The precise form of the 
 underlying CG gauge transformations in terms of the redefined fields~(\ref{field_redef}) is:
\bea\label{nonlin tr 2}
	&&
	\delta g_{\mu\nu}=\ {\cal L}_\xi g_{\mu\nu}\, 
	-\, \tfrac6\Lambda\,\big[\,\nabla_{\mu}\,\partial_{\nu}
	+\tfrac6\Lambda\,[(g+\tfrac6\Lambda\,\varphi)^{-1}]^{\rho\sigma}\,
	\gamma_{\rho\mu\nu}\,\partial_{\sigma}\,\big]\,\alpha\,,\\[2mm]
	&&
	\delta \varphi_{\mu\nu}=
	\mathcal L_{\xi}\,\varphi_{\mu\nu}
	+\big[\,\nabla_{\mu}\,\partial_{\nu}+\tfrac\Lambda3\,g_{\mu\nu}
	+\tfrac6\Lambda\,[(g+\tfrac6\Lambda\,\varphi)^{-1}]^{\rho\sigma}\,
	\gamma_{\rho\mu\nu}\,\partial_{\sigma}
	+2\,\varphi_{\mu\nu}\,\big]\,\alpha\, .\nonumber
\eea
Here we have denoted the Christoffel symbols
of~$\varphi_{\mu\nu}$, covariantized with respect to~$g_{\mu\nu}$, by
%\edz{A: The 1/2 in~$\xi=\frac 12 d \alpha$ is fishy and we  should have~$(grad^2 + \Lambda/3)\alpha$ for the~PM field....}
\be
	\gamma_{\rho\mu\nu}:=
	\tfrac12\left(\nabla_{\mu}\varphi_{\nu\rho}+\nabla_{\nu}\varphi_{\mu\rho}
	-\nabla_{\rho}\varphi_{\mu\nu}\right)\,.
\ee
Firstly observe  that at  leading order in~$\varphi$, the choice of diffeomorphism parameter~$\xi_\mu = 3\,\partial_\mu \alpha/\Lambda$ cancels the Lie derivative term
${\cal L}_\xi g_{\mu\nu} = \nabla_{\mu}\,\xi_{\nu}+\nabla_{\nu}\,\xi_{\mu}$ against the double gradient of the scalar parameter~$\alpha$ in the metric variation.
This is just a restatement of our linear result that the dynamical metric can be decoupled (at that order), leaving  linear~PM. Consistency of the
non-linear truncation requires that there exist a choice of~$\xi$ achieving this cancellation to all orders. This would determine the higher order terms in the variation of~$\varphi$, leaving the~PM action~$S_{\sss\rm PM}[\varphi]$ invariant. To establish a no-go result, we need only show that already 
no choice of~$\xi$ achieves this cancellation for the next-to-leading (linear) order terms in~$\varphi$ in the metric variation.
Focussing on the linear part~$\gamma^{\rho}_{\mu\nu}\partial_{\rho}\alpha$ of~$\delta g_{\mu\nu}$ we immediately see that it
can never be written as~$\nabla_{(\mu}X_{\nu)}$, for any~$X_{\nu}$ even on~PM-shell.
This establishes our claimed no-go result for truncating CG to a~PM theory beyond linear order.

Finally, we turn to the computation of the cubic  vertices. These, being  guaranteed  invariant under  leading~PM gauge transformations 
$\delta \varphi_{\mu\nu} = \big(\nabla_\mu\partial_\nu+\frac\Lambda 3g_{\mu\nu}\big) \alpha$,  are  candidate vertices for a putative non-linear self-inter\-acting PM theory.
The form of~$n$-th order Lagrangian of the~PM field determined by the field redefinition~\eqn{field_redef} (with~$J_{\mu\nu}=0=K_{\mu\nu}$) can be
obtained from the following correspondence,
%\bea\label{n-th vertex}
%	 (\tfrac\Lambda6)^{n+1}\sqrt{-g}\ \mathscr L^{(n+2)}_{\rm\sss~PM}
%	&=&\tfrac{n+1}{(n+2)!}\,\delta^{n+1}_{g|_\varphi}\!
%	\left[\sqrt{-g}\,\cal G^{\mu\nu} \right]
%	\varphi_{\mu\nu} \\[2mm]
%	&+&
%	\tfrac\Lambda6\,\tfrac1{n!}\,\delta^{n}_{g|_\varphi}\!\left[
%	\sqrt{-g}\,g^{\mu\nu}\,g^{\rho\sigma}\right]
%	\left(\varphi_{\mu\rho}\,\varphi_{\nu\sigma}-
%	\varphi_{\mu\nu}\,\varphi_{\rho\sigma}\right).\nn
%\eea
\bea\label{n-th vertex}
	 (\tfrac\Lambda6)^{n+1}\sqrt{-g}\ \mathscr L^{\sss (n+2)}_{\rm\sss~PM}
	&\!=\!&\tfrac{n+1}{(n+2)!}\,\varphi_{\mu\nu}\, \delta^{n+1}_{g|_\varphi}\!
	\left[\sqrt{-g}\, \cal G^{\mu\nu} \right]
	 \\[2mm]
	&&+\,
	\tfrac\Lambda6\,\tfrac1{n!}\delta^{n}_{g|_\varphi}\left[\sqrt{-g} \, g^{\mu\nu}g^{\rho\sigma}\right]\ 
        \left(\varphi_{\mu\rho}\,\varphi_{\nu\sigma}-
	\varphi_{\mu\nu}\,\varphi_{\rho\sigma}\right).\nn
\eea
Here~$\delta^{n}_{g|_\varphi}$ signifies taking the~$n$-th variation with respect to the {\it metric} and then replacing~$\delta g_{\mu\nu}$ by~$\varphi_{\mu\nu}$; the result is 
of~$n$-th order in~$\varphi_{\mu\nu}$. 
In the first line, we have used the fact that the first metric variation of the cosmological Einstein--Hilbert action produces the cosmological Einstein tensor~${\cal G}_{\mu\nu}$,
which allows~$(n+2)$ variations of that term to be combined with~$(n+1)$ variations of the coupling of the cosmological Einstein tensor to the partially massless field in~\eqn{kaku}.
If we evaluate the above interaction Lagrangians explicitly then, since
they are given in terms of multiple variations of the Ricci tensor,  the generic outcome for~$\mathscr L_{\sss\rm PM}$ is 
a two-derivative self-coupling of~$\varphi_{\mu\nu}$, a curvature coupling
and a potential for~$\varphi_{\mu\nu}$\,.

We also note that multiplying the original CG action~\eqn{W2action} by the dimension-free combination $\Lambda^{-1}\kappa^{-2}$ of the cosmological constant and gravitational couplings
and redefining the PM field $\varphi\to\Lambda\,\kappa\,  \varphi$ gives, schematically, the canonically normalized action $$S\sim \frac1{\kappa^2} \int (R-2\Lambda) +\int \Big[(\nabla \varphi)^2+\Lambda \varphi^2 \Big]+ \sum_{n=3}^\infty \kappa^{n-2} \big[\varphi^{n-2}\,\nabla \varphi\, \nabla \varphi  + \Lambda \varphi^{n}\big]\, .$$

Now, let us focus on computing the cubic part~$\mathscr L_{\rm \sss~PM}^{\sss (3)}$ in~(\ref{n-th vertex}).
Note that  since we work on an Einstein background, we may set  ${\cal G}_{\mu\nu}=0$ (when it is not varied); also, since we only quote the vertex up to a possible
field redefinition, at this order we may use the linear PM field equation, 
which can be written as
$
\delta_{g|_\varphi} {\mathcal G}_{\mu\nu}+\tfrac{\Lambda}{3} \big(\varphi_{\mu\nu} - g_{\mu\nu}\varphi\big)=0\,
$.\footnote{Notice that the cubic vertex, therefore,  schematically takes the form $$S^{\sss(3)}_{\sss\rm PM}=\delta_{g|_\varphi} S^{\sss(2)}_{\sss\rm PM}\,  +\, \int  \varphi^3\, ,$$ where $S^{\sss(2)}_{\sss\rm PM}$ is the leading order PM action and $\varphi^3$ 
denotes cubic potential terms in $\varphi_{\mu\nu}$.}  
Moreover, since the vertex is cubic in $\varphi_{\mu\nu}$, we may write 
$$
\tfrac6\Lambda\,  T^{\mu\nu}:= \frac13\,  \frac{1}{\sqrt{-g}} 
\frac{\delta S^{\sss (3)}_{\sss\rm PM}}{\delta \varphi_{\mu\nu}}\quad \mbox{and}\quad
S^{\sss (3)}_{\sss\rm PM}=\tfrac 6\Lambda\int \sqrt{-g} \ \varphi_{\mu\nu}\,  T^{\mu\nu} \, .
$$
By construction, $S^{\sss(3)}_{\sss\rm PM}$ is invariant under the linear order PM gauge transformation~\eqn{pm_gauge} modulo the linear field equations. This guarantees that $T_{\mu\nu}$ obeys Noether identity
\be\label{noether}
(\nabla^\mu\nabla^\nu +\tfrac \Lambda3\,g^{\mu\nu})\,
T_{\mu\nu} \approx 0\, ,
\ee
in an Einstein background
where $\approx$ denotes equality modulo the linear PM field equations.

It remains to explicitly compute $T_{\mu\nu}$. In fact, the cubic vertex itself is easily computed by hand by performing the 
variations of equation~\eqn{n-th vertex} for $n=1$. A computer aided computation~\cite{Vermaseren:2000nd} gives
\begin{eqnarray*}
T_{\mu\nu}&\!\approx\!&
\varphi^{\rho\sigma}\, \nabla_\rho\nabla_\sigma \varphi_{\mu\nu}+\tfrac12\, \varphi_{\mu\nu}\,\Delta\varphi
-\tfrac43\, \varphi^{\rho\sigma}\,\nabla_{(\mu|}\nabla_\rho\varphi_{\sigma|\nu)}-\varphi^\rho{}_{\!\!(\mu}\,\nabla_{\nu)}\nabla_\rho\varphi \\[1mm]
&&+\,\tfrac23\,\varphi^{\rho\sigma}\,\nabla_\mu\nabla_\nu \varphi_{\rho\sigma}
+\tfrac16\,\varphi\,\nabla_\mu\nabla_\nu \varphi
+\tfrac1{6}\, g_{\mu\nu}\left( \varphi^{\rho\sigma}\,\nabla_\rho\nabla_\sigma \varphi-\varphi\, \Delta\varphi\right)\\[1mm]
&&+\, \nabla^\rho \varphi\, (\tfrac 32\,  \nabla_\rho \varphi_{\mu\nu} -\tfrac 23 \, \nabla_{(\mu}\varphi_{\nu)\rho})
-\tfrac13\, \nabla^\rho \varphi^\sigma{}_{\!\!\mu}\, \nabla_\rho \varphi_{\sigma\nu}
- \nabla^\rho \varphi^\sigma{}_{\!\!(\mu|} \,\nabla_\sigma \varphi_{|\nu)\rho} \\[1mm]
&&+\,\tfrac23\, \nabla_{(\mu|}\varphi^{\rho\sigma}\, \nabla_{\rho|}\varphi_{\nu)\sigma}
+\tfrac16\, \nabla_{\mu}\varphi^{\rho\sigma}\, \nabla_{\nu}\varphi_{\rho\sigma}-\tfrac13 \, \nabla_\mu\varphi\, \nabla_\nu \varphi \\[1mm]
&&-\, g_{\mu\nu}\,(\tfrac5{12}\, \nabla^\rho\varphi^{\sigma\tau}\,\nabla_\rho\varphi_{\sigma\tau}
-\tfrac12\,  \nabla^\rho\varphi^{\sigma\tau}\,\nabla_\sigma\varphi_{\rho\tau}+\tfrac1{12}\, \nabla^\rho \varphi\, \nabla_\rho \varphi) \\[1mm]
&&-\, \Lambda \, (\tfrac1 {18}\, \varphi \, \varphi_{\mu\nu} + \tfrac{5}9\, \varphi^\rho{}_{\!\! \mu}\, \varphi_{\nu\rho})
+\Lambda\,  g_{\mu\nu}\, ( \tfrac{11}{36}\, \varphi^{\rho\sigma}\,\varphi_{\rho\sigma}-\tfrac1{36} \, \varphi^2 ) \\[1mm]
&&-\,\tfrac23 \, W^{\rho\tau}{}_{(\mu}{}^\sigma \varphi_{\nu)\tau}\,\varphi_{\rho\sigma}
-\tfrac23 \, W^\rho{}_{(\mu\nu)}{}^\sigma \varphi^\tau{}_{\!\! \rho}\,  \varphi_{\tau\sigma} 
-\tfrac13\, g_{\mu\nu}\, W^{\rho\tau\kappa\sigma}\varphi_{\rho\sigma}\,\varphi_{\tau\kappa}\, .
\end{eqnarray*}
As a check, we verified that this $T_{\mu\nu}$ obeys the Noether identity~\eqn{noether} for the case of
constant curvature $g_{\mu\nu}$.

 We emphasize again that 
the consistency of the cubic vertices is independent of that of the higher order ones,
implying that  the vertex~$S^{\sss(3)}_{\sss\rm PM}$ is valid beyond the context of CG.
In fact, this vertex has been constructed 
 in~\cite{Zinoviev:2006im} where it has been shown that two-derivative self-interactions of
~PM fields exist only for~$d=4$\,. We now see that CG underlies that result which
% Moreover, in that work, it has been argued that the aforementioned particularity of~$d=4$
% should be related to the fact that the~PM spin 2 is 
% conformally invariant only in~$d=4$~\cite{}. The derivation of these~PM self-interactions from CG clarifies this observation since
% its (non-linear) Weyl invariance is valid only in~$d=4$\,.
 is also consistent with the recent work of~\cite{Joung:2012rv} where all consistent cubic interactions 
 (not necessarily  two-derivative ones)
 involving PM fields of generic spin were considered. There 
  it was shown that for generic dimensions there are only two~PM self-couplings
  involving at most four and six derivatives respectively.
 However, precisely in four dimensions,  the Gau\ss-Bonnet identity reduces the maximal
 four-derivative coupling to a two-derivative one\footnote{In fact, for constant curvature backgrounds, the Cotton-like tensor~\cite{Deser:2006zx} 
 $$F_{\mu\nu}{}^\rho:=\nabla_{\mu} \varphi_{\nu}{}^{\!\rho} - 
 \nabla_{\nu}\varphi_\mu{}^{\!\rho}$$ is invariant under PM gauge transformations~\eqn{pm_gauge}. (Strictly this version of the Cotton tensor is not the metric one, because the PM field is not the Schouten tensor, although in the underlying CG setting this 
is  in fact the case.)
 Therefore any quartic derivative order, cubic vertex of type $\int (\nabla F) F F$ is PM invariant. In four dimensions, it should be possible to employ the Gau\ss--Bonnet identity to write this as a manifestly invariant cubic vertex quadratic in derivatives. }.

\section{Conclusions}

We have revisited the well-trodden grounds of~$d=4$ conformal, Weyl, gravity to explore the coupling of PM spin~2 fields to 
 both fixed nontrivial curved backgrounds and dynamical gravity, as well as their self-interactions.  
 The first result was that fixed Bach-flat backgrounds break the~PM gauge invariance, and no natural
background more general than Einstein was found in the space of all possible couplings to geometry and extensions of the gauge invariance for the linear PM theory.
For dynamical gravity couplings, the attraction of CG was that it---at linearized level about dS---is the ``difference'' between a normal graviton and a~PM one. 
However, upon insisting on a  ghost-free model one must
% of two novel possibilities for the~PM field, 
give up on any dynamical pretensions for the (ghost) graviton.
%But, since the theory is fourth derivative order, those two components---the result of an Ostrogradski auxiliary field decomposition---are necessarily of relative ghost sign, and so physically unacceptable. 
%Apart from this ``minor'' unitarity requirement, Weyl (or indeed any other quadratic gravity model but pure~$R^2$) would indeed exemplify a consistent fully non-linear tensor-dynamic gravity realization. 
%The first, more modest one, keeps to linear~PM and explores its viability in a wider class of backgrounds than Einstein ones, as naturally suggested by polarizing Weyl into its separate components and testing for logical extended backgrounds. %That is, while it is not surprising that a free tensor field, of any rank [ we 0408155]?), is consistent in~dS as well as flat geometries, we tested for its survival beyond Einstein spaces, and indeed found it didn't.
This left the possibility that (just as truncating CG to its PM vacuum yields cosmological Einstein gravity) an appropriate truncation of the dynamical graviton could 
allows CG to generate consistent PM self-interactions.
%Our second, more ambitious, search was for a consistent, self-interacting--non-linear--extension of~PM, if still on Einstein backgrounds: giving up some geometric generality for consistent non-linearity. 
Here too we came up against  the old problem, one that already occurs in trying to similarly extend higher-spin gauge theories: %(besides GR itself, of course) in flat or~dS spaces
while it is  possible to set up a---lowest order invariant---cubic self-interaction expressed as the coupling of the (quadratic) Noether current maintaining the initial Abelian invariance to the field amplitude, self-coupling inconsistencies set in at quartic order. That is, within the CG setting, one cannot extend the initial Abelian invariance to cover the next non-linear vertex. We illustrated this by exhibiting both the corresponding cubic terms and showing that there is no combination of fields, inert under (non-linear) PM transformations to the next order, truncating to  consistent PM in a gravitational background. Although CG underlies cosmological Einstein gravity, it does not truncate to a non-linear ``PM general relativity''.  

No-go theorems are notorious for their loopholes. Spin (2,3/2)-gravity
and supergravity theories circumvent just such higher-spin pitfalls~\cite{Buchdahl} while for (towers of) massive higher spins, string theory provides presumably consistent interactions and Vasiliev's 
theory describes interactions of towers of massless higher spins in (A)dS backgrounds~\cite{Vasiliev}. 
Nonetheless,  our results relying on CG as the underpinning of~PM self-interactions seem quite robust. As stated in~\cite{Zinoviev:2006im},
``We have checked that it is impossible to proceed with quadratic approximation without introduction of higher derivative terms and/or some other fields.'' 
Moreover, also there, only a single~PM cubic vertex was found; its agreement with the one generated by CG would suggest the absence of a self-interacting, two-derivative~PM theory is a likely outcome.

One interesting feature of CG is that the~PM field can be consistently turned off, leaving cosmological Einstein gravity (at least classically). 
In other words, without additional matter couplings, 
choosing  initial  conditions such that~$\varphi_{\mu\nu}$ is zero at the initial time, then
it will remain  trivial, while the metric~$g_{\mu\nu}$ can realize 
any  Einstein solution~\cite{Maldacena:2011mk}. We could also envisage a situation where the~PM field~$\varphi_{\mu\nu}$ is not strictly zero but rather  nearly zero in
some arbitrarily large time interval~$t_{i}\ldots t_{f}$.
Cosmology would then have approximate Einstein behavior for that epoch, while in the region~$t  \ll t_{i}$ or~$t \gg t_{f}$, non-Einstein solutions could emerge.
(The consequences for cosmological expansion with a partially conserved symmetric two index boundary operator were also considered in~\cite{Dolan:2001ih}.) CG~could then be used to generate transitions from a~dS inflationary behavior of the cosmic scale factor to  one controlled by the partially massless modes.
Ghosts and loss of stability at early and late times may even be a useful/acceptable feature in this scenario.

A separate speculation is that gravity-like, or even self-interacting PM-like models for higher~$s>2$, might be achievable by studying higher-spin 
versions of CG. Indeed, interacting conformally invariant higher-spin models that can be viewed as analogs of CG do exist~\cite{Segal:2002gd, Bekaert:2010ky}. Perhaps a higher spin version of our approach could apply there.

\section*{Acknowledgements}
We thank Hamid Afshar, Rod Gover, Daniel Grumiller and Karapet Mkrtch\-yan for fruitful discussions.
EJ and AW acknowledge the ESI Vienna Workshop on Higher Spin Gravity. 
SD was supported in part by NSF PHY-1064302 and DOE DE-FG02-164 92ER40701 grants.
The work of EJ was supported in part by Scuola Normale Superiore, by INFN (I.S. TV12) and by the MIUR-PRIN contract 2009-KHZKRX.


\begin{thebibliography}{99} 

%\bibitem{Weyl} H.~Weyl, %``Gravitation und Elektrizit\"at'', Sitz. Preuss. Akad. Wiss. 465, (1918).
%{\tt \small Cited on page(s) \hspace{-3mm}}

%\cite{Deser:1983tm}
\bibitem{Deser:1983tm}
S.~Deser and R.~I.~Nepomechie,
%``Anomalous Propagation Of Gauge Fields In Conformally Flat Spaces,''
Phys.\ Lett.\ B {\bf 132}, 321 (1983);
%%CITATION = PHLTA,B132,321;%%
%\cite{Deser:1984mm}
%\bibitem{Deser:1984mm}
%S.~Deser and R.~I.~Nepomechie,
%``Gauge Invariance Versus Masslessness In De Sitter Space,''
Annals Phys.\ {\bf 154}, 396 (1984). {\tt \small Cited on pages \hspace{-3mm}}
%%CITATION = APNYA,154,396;%%

%\cite{Deser:2001pe}
\bibitem{Deser:2001pe}
S.~Deser and A.~Waldron,
%``Gauge invariances and phases of massive higher spins in (A)dS,''
Phys.\ Rev.\ Lett.\  {\bf 87}, 031601 (2001)
[arXiv:hep-th/0102166];
%%CITATION = HEP-TH 0102166;%%
%\cite{Deser:2001us}
%\bibitem{Deser:2001us}
%S.~Deser and A.~Waldron,
%``Partial masslessness of higher spins in (A)dS,''
Nucl.\ Phys.\ B {\bf 607}, 577 (2001)
[arXiv:hep-th/0103198].
%%CITATION = HEP-TH 0103198;%%
{\tt \small Cited on page \hspace{-3mm}}

%\cite{Higuchi:1986py}
\bibitem{Higuchi:1986py} 
  A.~Higuchi,
  %``Forbidden Mass Range For Spin-2 Field Theory In De Sitter Space-time,''
  Nucl.\ Phys.\ B {\bf 282}, 397 (1987);
  %%CITATION = NUPHA,B282,397;%%
  %\cite{Higuchi:1989gz}
%\bibitem{Higuchi:1989gz} 
  %A.~Higuchi,
  %``Massive Symmetric Tensor Field In Space-times With A Positive Cosmological Constant,''
  Nucl.\ Phys.\ B {\bf 325}, 745 (1989);
  %%CITATION = NUPHA,B325,745;%%
  %\cite{Higuchi:1986wu}
%\bibitem{Higuchi:1986wu} 
  %A.~Higuchi,
  %``Symmetric Tensor Spherical Harmonics On The N Sphere And Their Application To The De Sitter Group So(n,1),''
  J.\ Math.\ Phys.\  {\bf 28}, 1553 (1987)
  [Erratum-ibid.\  {\bf 43}, 6385 (2002)].
  %%CITATION = JMAPA,28,1553;%%
{\tt \small Cited on page \hspace{-3mm}}



%\cite{Deser:2001wx}
\bibitem{Deser:2001wx}
S.~Deser and A.~Waldron,
%%``Stability of massive cosmological gravitons,''
Phys.\ Lett.\ B {\bf 508}, 347 (2001)
[arXiv:hep-th/0103255].
%%CITATION = HEP-TH 0103255;%%
{\tt \small Cited on page \hspace{-3mm}}

%\cite{Deser:2001xr}
\bibitem{Deser:2001xr}
S.~Deser and A.~Waldron,
%``Null propagation of partially massless higher spins in (A)dS and
%cosmological constant speculations,''
Phys.\ Lett.\ B {\bf 513}, 137 (2001)
[arXiv:hep-th/0105181].
%%CITATION = HEP-TH 0105181;%%
{\tt \small Cited on page \hspace{-3mm}}


%\cite{Deser:2003gw}
\bibitem{Deser:2003gw}
S.~Deser and A.~Waldron,
%%``Arbitrary spin representations in de Sitter from~dS/CFT with applications to
%dS supergravity,''
Nucl.\ Phys.\ B {\bf 662}, 379 (2003)
[arXiv:hep-th/0301068].
%%CITATION = HEP-TH 0301068;%%
{\tt \small Cited on page \hspace{-3mm}}


%\cite{Maldacena:2011mk}
\bibitem{Maldacena:2011mk} 
  J.~Maldacena,
  %``Einstein Gravity from Conformal Gravity,''
  arXiv:1105.5632 [hep-th].
  %%CITATION = ARXIV:1105.5632;%% 
  {\tt \small Cited on pages \hspace{-3mm}}

%\cite{Dolan:2001ih}
\bibitem{Dolan:2001ih} 
  L.~Dolan, C.~R.~Nappi and E.~Witten,
  %``Conformal operators for partially massless states,''
  JHEP {\bf 0110}, 016 (2001)
  [hep-th/0109096].
  %%CITATION = HEP-TH/0109096;%%
{\tt \small Cited on pages \hspace{-3mm}}


%\cite{Deser:2004ji}
\bibitem{Deser:2004ji}
  S.~Deser and A.~Waldron,
  %``Conformal invariance of partially massless higher spins,''
  Phys.\ Lett.\ B {\bf 603}, 30 (2004),
  arXiv:hep-th/0408155.
  %%CITATION = HEP-TH 0408155;%%
{\tt \small Cited on pages \hspace{-3mm}}






%\cite{Deser:2006zx}
\bibitem{Deser:2006zx} 
  S.~Deser and A.~Waldron,
  %``Partially Massless Spin 2 Electrodynamics,''
  Phys.\ Rev.\ D {\bf 74}, 084036 (2006)
  [hep-th/0609113].
  %%CITATION = HEP-TH/0609113;%%
{\tt \small Cited on pages \hspace{-3mm}}


%\cite{Zinoviev:2009hu}
\bibitem{Zinoviev:2009hu} 
  Y.~M.~Zinoviev,
  %``On massive spin 2 electromagnetic interactions,''
  Nucl.\ Phys.\ B {\bf 821}, 431 (2009)
  [arXiv:0901.3462 [hep-th]].
  %%CITATION = ARXIV:0901.3462;%%
{\tt \small Cited on page \hspace{-3mm}}




%\cite{Zinoviev:2006im}
\bibitem{Zinoviev:2006im}
  Y.~.M.~Zinoviev,
  %``On massive spin 2 interactions,''
  Nucl.\ Phys.\ B {\bf 770} (2007) 83
  [hep-th/0609170].
  %%CITATION = HEP-TH/0609170;%%
{\tt \small Cited on pages \hspace{-3mm}}

%\cite{Joung:2012rv}
\bibitem{Joung:2012rv} 
  E.~Joung, L.~Lopez and M.~Taronna,
  %``On the cubic interactions of massive and partially-massless higher spins in (A)dS,''
  arXiv:1203.6578 [hep-th].
  %%CITATION = ARXIV:1203.6578;%%
{\tt \small Cited on pages \hspace{-3mm}}

%%\cite{deRham:2010kj}
%\bibitem{deRham:2010kj} 
%  C.~de Rham, G.~Gabadadze and A.~J.~Tolley,
%  %``Resummation of Massive Gravity,''
%  Phys.\ Rev.\ Lett.\  {\bf 106}, 231101 (2011)
%  [arXiv:1011.1232 [hep-th]].
%  %%CITATION = ARXIV:1011.1232;%%
%{\tt \small Cited on page \hspace{-3mm}}

%\cite{deRham:2012kf}
\bibitem{deRham:2012kf} 
  C.~de Rham and S.~Renaux-Petel,
  %``Massive Gravity on de Sitter and Unique Candidate for Partially Massless Gravity,''
  arXiv:1206.3482 [hep-th].
  %%CITATION = ARXIV:1206.3482;%%
  {\tt \small Cited on page \hspace{-3mm}}
  
\bibitem{2+2+2}
%\cite{Stelle:1976gc}
%\bibitem{Stelle:1976gc} 
  K.~S.~Stelle,
  %``Renormalization of Higher Derivative Quantum Gravity,''
  Phys.\ Rev.\ D {\bf 16}, 953 (1977);
  %%CITATION = PHRVA,D16,953;%%
  %\cite{Ferrara:1977mv}
%\bibitem{Ferrara:1977mv} 
  S.~Ferrara and B.~Zumino,
  %%``Structure of Conformal Supergravity,''
  Nucl.\ Phys.\ B {\bf 134}, 301 (1978);
  %%CITATION = NUPHA,B134,301;%%
E.~S.~Fradkin and A.~A.~Tseytlin,
 %``Higher Derivative Quantum Gravity: One Loop Counterterms and
%Asymptotic Freedom,''
 Nucl.\ Phys.\ B {\bf 201}, 469 (1982);
 %\cite{Fradkin:1985am}
%\bibitem{Fradkin:1985am} 
 % E.~S.~Fradkin and A.~A.~Tseytlin,
  %%``Conformal Supergravity,''
  Phys.\ Rept.\  {\bf 119}, 233 (1985);
  %%CITATION = PRPLC,119,233;%%
  %\cite{Riegert:1984hf}
%\bibitem{Riegert:1984hf} 
  R.~J.~Riegert,
  %%``The Particle Content Of Linearized Conformal Gravity,''
  Phys.\ Lett.\ A {\bf 105}, 110 (1984).
  %%CITATION = PHLTA,A105,110;%%
 {\tt \small Cited on page \hspace{-3mm}}  

%\cite{Lu:2011ks}
\bibitem{Lu:2011ks} 
  H.~Lu, Y.~Pang and C.~N.~Pope,
  %``Conformal Gravity and Extensions of Critical Gravity,''
  Phys.\ Rev.\ D {\bf 84}, 064001 (2011)
  [arXiv:1106.4657 [hep-th]].
  %%CITATION = ARXIV:1106.4657;%%    
{\tt \small Cited on page \hspace{-3mm}}  
  
%\cite{Metsaev:2007fq}
\bibitem{Metsaev:2007fq} 
  R.~R.~Metsaev,
  %``Ordinary-derivative formulation of conformal low spin fields,''
  JHEP {\bf 1201}, 064 (2012)
  [arXiv:0707.4437 [hep-th]].
  %%CITATION = ARXIV:0707.4437;%%  
  {\tt \small Cited on page \hspace{-3mm}}
  
  
%\cite{Lu:2011zk}
\bibitem{Lu:2011zk} 
  H.~Lu and C.~N.~Pope,
  %``Critical Gravity in Four Dimensions,''
  Phys.\ Rev.\ Lett.\  {\bf 106}, 181302 (2011)
  [arXiv:1101.1971 [hep-th]].
  %%CITATION = ARXIV:1101.1971;%%
 {\tt \small Cited on page \hspace{-3mm}} 
  
  
%\cite{Hyun:2011ej}
\bibitem{Hyun:2011ej} 
  S.~-J.~Hyun, W.~-J.~Jang, J.~-H.~Jeong and S.~-H.~Yi,
  %``Noncritical Einstein-Weyl Gravity and the AdS/CFT Correspondence,''
  JHEP {\bf 1201}, 054 (2012)
  [arXiv:1111.1175 [hep-th]].
  %%CITATION = ARXIV:1111.1175;%%
{\tt \small Cited on page \hspace{-3mm}}
  
  
  %\cite{Kaku:1977pa}
\bibitem{Kaku:1977pa} 
  M.~Kaku, P.~K.~Townsend and P.~van Nieuwenhuizen,
  %``Gauge Theory of the Conformal and Superconformal Group,''
  Phys.\ Lett.\ B {\bf 69}, 304 (1977).
  %%CITATION = PHLTA,B69,304;%%
  {\tt \small Cited on pages \hspace{-3mm}}
  
  \bibitem{BEG} T.N.\ Bailey, M.G.\ Eastwood, and A.R.\ Gover,
   %``Thomas's structure bundle for conformal, projective and related structures'', 
   Rocky Mountain J.\ Math.\ {\bf 24} (1994),
  1191.
  %\bibitem{Goinvariant}
A.R.~Gover, 
%``Invariant theory and calculus for conformal geometries'', 
Adv. Math. {\bf 163} (2001), 206--257.
{\tt \small Cited on pages \hspace{-3mm}}

\bibitem{Goal} A.R. Gover, 
   %``Almost Einstein and Poincar\'e-Einstein manifolds in Riemannian signature'',   
    J.\ Geometry and 
   Physics, {\bf 60} (2010), 182,   arXiv:0803.3510.
   {\tt \small Cited on page \hspace{-3mm}}

\bibitem{TMG} S.~Deser, R.~Jackiw and  S. Templeton, Ann Phys {\bf 140}, (1982) 372; Phys. Rev. Lett. {\bf 48}, (1982) 975.
{\tt \small Cited on page \hspace{-3mm}}    
    
%\cite{Gover:2006fa}
\bibitem{Gover:2006fa} 
  A.~R.~Gover, P.~Somberg and V.~Soucek,
  %``Yang-Mills detour complexes and conformal geometry,''
  Commun.\ Math.\ Phys.\  {\bf 278}, 307 (2008)
  [math/0606401 [math-dg]].
  %%CITATION = MATH/0606401;%%    
 {\tt \small Cited on pages \hspace{-3mm}}   
   
%\cite{Vermaseren:2000nd}
\bibitem{Vermaseren:2000nd}
  J.~A.~M.~Vermaseren,
  ``New features of FORM,''
  [arXiv:math-ph/0010025].
  %%CITATION = MATH-PH 0010025;%%
{\tt \small Cited on page \hspace{-3mm}}
  
 \bibitem{Buchdahl}
 H.~ Buchdahl,  Nuovo Cim.\  {\bf 10}, 96 (1958). 
{\tt \small Cited on page \hspace{-3mm}}  
  
\bibitem{Vasiliev}   
  M.~A.~Vasiliev,
  %``Consistent equation for interacting gauge fields of all spins in (3+1)-dimensions,''
  Phys.\ Lett.\ B {\bf 243}, 378 (1990).
  %%CITATION = PHLTA,B243,378;%%  
   {\tt \small Cited on page \hspace{-3mm}}   
   
   
%\cite{Segal:2002gd}

\bibitem{Segal:2002gd} 
  A.~Y.~Segal,
  %``Conformal higher spin theory,''
  Nucl.\ Phys.\ B {\bf 664}, 59 (2003)
  [hep-th/0207212].
  %%CITATION = HEP-TH/0207212;%%
  {\tt \small Cited on page \hspace{-3mm}}
 
 %\cite{Bekaert:2010ky}
\bibitem{Bekaert:2010ky} 
  X.~Bekaert, E.~Joung and J.~Mourad,
  %``Effective action in a higher-spin background,''
  JHEP {\bf 1102}, 048 (2011)
  [arXiv:1012.2103 [hep-th]].
  %%CITATION = ARXIV:1012.2103;%%
   {\tt \small Cited on page \hspace{-3mm}}

 

%%%%%%%%%%%%%%%%%%%















\end{thebibliography}
\end{document}